# Absence of Colossal Magnetoresistance in the Oxypnictide PrMnAsO$_{0.95}$F$_{0.05}$

Eve. J. Wildman [1], Falak Sher [2] and Abbie. C. Mclaughlin* [1]

[1] The Chemistry Department, University of Aberdeen, Meston Walk, Aberdeen, AB24 3UE, Scotland.

[2] Department of Chemistry, SBA School of Science and Engineering, LUMS, Lahore, Pakistan

**Abstract**

We have recently reported a new mechanism of colossal magnetoresistance in electron doped Mn oxypnictides NdMnAsO$_{1-x}$F$_x$. Magnetoresistances of up to -95 % at 3 K have been observed. Here we show that upon replacing Nd for Pr, the CMR is surprisingly no longer present. Instead a sizeable negative magnetoresistance is observed for PrMnAsO$_{0.95}$F$_{0.05}$ below 35 K (MR$_{7T}$ (12 K) = -13.4 % for PrMnAsO$_{0.9}$F$_{0.05}$). A detailed neutron and synchrotron X-ray diffraction study of PrMnAsO$_{0.95}$F$_{0.05}$ has been performed, which shows that a structural transition, $T_s$, occurs at 35 K from tetragonal *P*4*/nmm* to orthorhombic *Pmmn* symmetry. The structural transition is driven by the Pr 4f electrons degrees of freedom. The sizeable –MR observed below the transition most likely arises due to a reduction in magnetic and/or multipolar scattering upon application of a magnetic field.

## Introduction

Due to the recent discovery of high temperature superconductivity at 26 K in the electron doped 1111 iron pnictide *Ln*FeAsO [1] there has been much active research into oxypnictide materials. [2] Superconducting transition temperatures ($T_c$) of ≤ 56.3 K have been achieved via the substitution of oxygen with fluorine [3, 4, 5], the creation of oxygen vacancies [6] or by $Th^{4+}$ substitution for $Ln^{3+}$. [7] These materials form with the primitive tetragonal ZrCuSiAs structure (space group P4/*nmm*) and various other transition metal analogues have been reported, for example *Ln*MAsO (*Ln* = lanthanide, M = Mn, Co, Ni). [8, 9, 10] Recently, we reported colossal magnetoresistance (CMR) at low temperature in the antiferromagnetic oxypnictide series $NdMnAsO_{1-x}F_x$ ($x$ = 0 – 0.08). [11] CMR is a rare phenomenon [12, 13] which is well known in other manganese oxides such as the perovskite $Ln_{1-x}A_xMnO_3$ (*Ln* = La, Pr and *A* = Sr, Ca) [14, 15] and pyrochlore $Tl_2Mn_2O_7$. [16] Magnetoresistance (MR) is defined as the change of electrical resistivity $\rho$ in an applied magnetic field H, so that MR = ($\rho$(H)-$\rho$(0))/$\rho$(0). Magnetoresistant materials are important for magnetic memory device applications.

The parent compound NdMnAsO exhibits several magnetic transitions. At 359 K the $Mn^{2+}$ spins align antiferromagnetically with moments aligned parallel to *c*. At 23 K ($T_{Nd}$) the $Nd^{3+}$ spins order antiferromagnetically with their moments aligned parallel to the basal plane. [8, 17] At the same time a spin reorientation of the Mn spins occurs; the Mn spins rotate from their previous alignment along the *c* axis to along *a* over a 3 K temperature interval. Doping with a small amount of fluorine to obtain phases of $NdMnAsO_{1-x}F_x$ drastically changes the electronic properties of the parent compound, but little change to the magnetic properties upon substitution were observed. Variable temperature powder neutron diffraction data showed antiferromagnetic order of the $Mn^{2+}$ moments below 356(2) K, with the same magnetic structure as previously reported for NdMnAsO [8, 18] so there is no appreciable change in $T_{Mn}$. Below 23 K ($T_{Nd}$) antiferromagnetic alignment of $Nd^{3+}$ spins along *a* was observed with a corresponding spin reorientation of $Mn^{2+}$ into the basal plane at 20K ($T_{SR}$), as described for NdMnAsO. [17] MR is observed below ~75 K and is greatly enhanced upon cooling below $T_{SR}$, so that CMR is observed at low temperature ($MR_{9T}$(3K)) = -95 %). Electron doping manganese pnictides with H⁻ has also been achieved with large negative magnetoresistance observed in the series LaMnAsO$_{1-x}$H$_x$. The maximum MR is observed in the sample $x = 0.08$ (MR$_{5T}$ (8 K) = -63 %), as the antiferromagnetic order of Mn spins is suppressed by the emergence of a ferromagnetic metallic phase.[19] Electron doped materials SmMnAsO$_{1-x}$ have also been prepared[20] and surprisingly a large positive magnetoresistance is observed at low temperature in metallic materials (MR ~ 60% at 2 K for $x = 0.3$). This effect has been attributed to a field induced change of the complex Fermi surface. Given the diverse MR properties observed so far for the electron doped 1111 Mn oxyarsenides (Ln = La, Nd, Sm) we have synthesised and investigated the structural, electronic and magnetic properties of PrMnAsO$_{0.95}$F$_{0.05}$.

**Experimental**

A polycrystalline sample of PrMnAsO$_{0.95}$F$_{0.05}$ and NdMnAsO$_{1-x}$F$_x$ ($x$ = 0.02, 0.05) were synthesised *via* a two-step solid-state reaction method. Initially, the LnAs (Ln = Pr, Nd) precursor was obtained by the reaction of Ln pieces (Aldrich 99.9%) and As (Alfa Aesar 99.999%) at 900°C for 24h in an evacuated, sealed quartz tube. The resulting precursor was then reacted with stoichiometric amounts of MnO$_2$, Mn and MnF$_2$ (Aldrich 99.99%), all powders were ground in an inert atmosphere and pressed into a pellet of 10mm diameter. The pellet was placed into a Ta crucible and sintered at 1150°C for 48h, again in a quartz tube sealed under vacuum.

Powder X-ray diffraction patterns of PrMnAsO$_{0.95}$F$_{0.05}$ and NdMnAsO$_{1-x}$F$_x$ ($x$ = 0.02, 0.05) were collected using a Bruker D8 Advance diffractometer with twin Gobel mirrors and Cu Kα radiation. Data were collected at room temperature over the range 10° < 2θ < 100°, with a step size of 0.02°, and could be indexed on a tetragonal unit cell of space group *P*4/*nmm*, characteristic of the ZrCuSiAs structure type as previously reported for PrMnAsO [21]. X-ray diffraction patterns demonstrated that the material was of high purity.

High resolution synchrotron X-ray powder diffraction patterns were recorded on the ID31 beamline for PrMnAsO$_{0.95}$F$_{0.05}$ at the ESRF, Grenoble, France at several temperatures between 10 K and 290 K with a wavelength of 0.3999 Å and a collection time of one hour at each temperature. The powder sample was inserted into a 0.5 mm diameter borosilicate glass capillary and spun at ~1Hz. The patterns were collected between $2^o < 2\theta < 50^o$.

Powder neutron diffraction data were also recorded for $PrMnAsO_{0.95}F_{0.05}$ using the D20 high intensity diffractometer at the Institute Laue Langevin (ILL, Grenoble, France). Neutrons of wavelength 2.4188 Å were incident on an 8mm vanadium can and data were recorded between 2 - 300 K in a cryostat and between 300 K – 360 K in a furnace, with a collection time of 20 minutes at each temperature. The temperature dependency of the magnetic structure was obtained by Rietveld refinement [22] of the neutron data using the GSAS package. [23]

The temperature and field dependence of the electrical resistance were recorded using a Quantum Design physical property measurement system (PPMS) between 4 and 300 K in magnetic fields of 7 T. The magnetic susceptibility was measured with a Quantum Design superconducting quantum interference device magnetometer (SQUID). Zero field cooled (ZFC) measurements were recorded between 2 and 400 K in a field of 1000 Oe.

## Results

**Crystal Structure**

The Rietveld refinement of high resolution X-ray powder diffraction data collected between 10 and 290 K confirmed that $PrMnAsO_{0.95}F_{0.05}$ crystallises at room temperature with the expected ZrCuSiAs-type tetragonal structure of space group $P4/nmm$ (Fig. 1). Similar to PrMnSbO [24], a structural transition to orthorhombic $Pmmn$ symmetry is observed below $T_s$ ~ 35 K, demonstrated by a subtle splitting of the (302), (310) and (311) reflections upon cooling as shown in the inset of Figure 1. $Pr^{3+}$ has the $4f^2$ configuration and the structural distortion is a result of multipolar order, most likely ferromultipolar order as previously reported for PrMnSbO [24]. In contrast, there is no evidence of a structural transition in $NdMnAsO_{1-x}F_x$ which remains in the $P4/nmm$ space group down to 4 K. [11]

Below $T_s \sim 35$ K the data were fit well with an orthorhombic unit cell of subgroup *Pmmn* ($a = 4.05896(1)$ Å, $b = 4.06201(1)$ Å and $c = 8.89399(2)$ Å at 10 K) where the Pr site symmetry undergoes a reduction from 4*mm* to *mm*2. This structural change differs from the orthorhombic distortion observed in the Fe analogue, LaFeAsO, which results in an enlarged centred cell of space group *Cmma*. The refined values for lattice constants, atomic parameters, selected bond lengths and angles with corresponding agreement indices for the respective variable temperature fits to the data are found in Table 1. There is no evidence of cation or As/O anion disorder. The Pr, Mn and As occupancies refined to within ± 1 % of the full occupancy and were fixed at 1.0. The O and F occupancies were fixed at 0.95 and 0.05 respectively.

The temperature dependence of the cell parameters are shown in Figure 2. The orthorhombic to tetragonal transition is clearly observed below $T_s \sim 35$ K. The structural transition is driven by Pr 4f electron degrees of freedom [24] which results in the splitting of the Pr-O bond lengths. At 35 K the Pr-O bond length is 2.3429(1) Å and splits into two bond lengths of 2.385(5) Å and 2.303(4) Å below $T_s$ (Fig. 3). In comparison a much smaller change is evident in the Mn-As bond lengths. There is also a contraction in both of the Pr-O-Pr and As-Mn-As blocks below this temperature (Table 1).

The variation of *c* with temperature is shown in the inset of Figure 2 and a subtle change in slope is apparent at ~ 140 K. There is also a discontinuity present in *a* at this temperature (Fig. 2). The Mn-As and Pr-O/F bond lengths and angles are shown in Figure 3 and Table 1 respectively. Subtle anomalies are observed in all of the Mn-As bond lengths and As-Mn-As angles at ~ 140 K. In contrast, there is no evidence of a discontinuity in any of the Pr-O bond lengths or Pr-O-Pr bond angles.

**Magnetic Structure**

Powder neutron diffraction measurements were recorded over the temperature range 2 - 360 K. Below $T_{Mn} = 340$ K the (100) magnetic reflection is observed alongside a magnetic contribution to the intensity of the (101) and (102) structural peaks as a result of antiferromagnetic ordering of the Mn Spins (Fig. 4). This is comparable to the Nd analogue (NdMnAsO$_{0.95}$F$_{0.05}$) with a $T_{Mn}$ of 356 K [11]. Above 340 K the (100) magnetic reflection is no longer observable but diffuse scattering characteristic of short range ordering is still present as shown in Figure 4. The magnetic reflections observed below 340 K could be indexed with the propagation vector ($k = 0, 0, 0$), so that the nuclear and magnetic unit cells are equivalent. Rietveld refinement of the data revealed that at $T_{Mn}$ the Mn moments align antiferromagnetically in the *ab* plane and ferromagnetically along *c* as previously reported for NdMnAsO$_{1-x}$F$_x$ [8, 11, 17] with spins parallel to the *c* axis.

The variation of the high spin Mn moment with temperature is displayed in Figure 5. Below ~ 220 K ($T_{SR1}$) the Mn spins start to reorient to align parallel to *a*, which results in a slight reduction in intensity of the (100) and (101) peaks but an increase in the intensity of the (102) reflection. There is no change in spin state of the Mn ion at the spin reorientation transition. At 180 K a poorer Rietveld fit to the (100) magnetic reflection is observed which can be modelled by adding a small antiferromagnetic moment on the Pr$^{3+}$ site. The Pr$^{3+}$ spins are aligned parallel to *a* (Fig. 4) and the concave variation of the Pr moment with temperature (Fig. 5) suggests that the magnetic ordering of the Pr$^{3+}$ spins is induced upon reorientation of the Mn spins into the basal plane. The spin reorientation of the Mn moments occurs over an ~ 80 K temperature interval so that by 140 K ($T_{SR2}$) the Mn spins are fully aligned parallel to *a*, with a moment of 3.17 (3) µ$_B$. The same spin reorientation is observed below 90 K in PrMnSbO over a 20 K temperature interval. [24] For PrMnSbO, the induced Pr antiferromagnetic order is only observed once the Mn moments are fully aligned in the basal plane. In contrast, in PrMnAsO$_{0.95}$F$_{0.05}$, the induced Pr moment is detected after the Mn spins start reorienting into the basal plane. It's most likely that a small Pr moment is actually also induced at 200 K but the magnetic intensity is too weak to be seen in our measurements. Similar behaviour is also observed in La$_{1-x}$Ce$_x$MnAsO [25] where spin reorientation of the Mn moments induces long range magnetic order of the Ce$^{3+}$ moments below 34 K. A simultaneous spin reorientation of the Mn moments and alignment of the Nd spins is also observed in NdMnAsO$_{1-x}$F$_x$ at the lower temperature of 23 K. [8, 11, 17]

Magnetic susceptibility measurements of PrMnAsO$_{0.95}$F$_{0.05}$ show an anomaly in the inverse susceptibility at T$_{Mn}$ ~ 340 K and a broad transition around 38 K (Fig. 6). There is no evidence of a magnetic transition in the neutron diffraction data at this temperature, so that the broad peak most likely arises from the crystalline field splitting of the Pr$^{3+}$ ion.

The high resolution synchrotron X-ray diffraction data identified a structural transition at $T_s$ ~ 35 K from tetragonal to orthorhombic symmetry, due to a small distortion in the *ab* plane (Figs 1 and 2). As $a \neq b$ below this temperature the neutron data were separately modelled with the moments along the *a* and *b* axes and then compared. The best results were obtained with magnetic moments ordered parallel to *b* ($\chi^2$ = 13.3 for moments aligned along *b* compared to 14.1 for moments aligned along *a* at 2 K); this arrangement was therefore adopted for all temperatures below $T_s$. The synchrotron data also evidenced anomalies in the cell parameters, Mn-As bond lengths and angles at ~ 140 K (Figs 2 and 3, Table 1), which corresponds to $T_{SR2}$, the temperature at which the Mn moments have fully re-orientated into the basal plane. This demonstrates a coupling between the crystal and magnetic lattices, i.e., the crystal lattice is sensitive to the orientation of the Mn moment. In corroboration there is no evidence of a discontinuity in any of the Pr-O/F bond lengths or Pr-O-Pr bond angles at 140 K.

**Electronic Properties**

PrMnAsO$_{0.95}$F$_{0.05}$ is semiconducting with a room temperature resistivity of $\rho_{290K}$ = 0.304 Ω.cm. Arrhenius behaviour is observed between 95 K – 220 K (Figure 7, inset b), where the resistivity, $\rho$, follows the relationship $\rho = \rho_0 \exp(E_g/2kT)$ (where $\rho$ is the measured resistivity, $E_g$ the band gap, $k$ the Boltzmann constant and $T$ is the temperature), with a calculated $E_g$ of 0.048(1) eV. The electron transport properties in this region are dominated by thermally activated charge carriers across a band gap. Similar results are obtained for NdMnAsO$_{0.95}$F$_{0.05}$ [11] where $E_g$ = 0.023(1) eV.

The low temperature resistivity of PrMnAsO$_{0.95}$F$_{0.05}$ is shown in Figure 7, inset a. Upon cooling PrMnAsO$_{0.95}$F$_{0.05}$ below 95 K, the temperature variation of the resistivity data is characteristic of Mott three-dimensional variable-range hopping (VRH) of the electrons (Fig. 7).[26] Variable range hopping conductivity implies that the electronic states are localised at the Fermi level by disorder. Disorder in a solid can introduce random potential energy in the lattice which can lead to localisation of the electronic wavefunction (Anderson localisation). Anderson localisation is caused by a disordered potential which could arise from the substitution of F$^-$ for O$^{2-}$ in PrMnAsO$_{0.95}$F$_{0.05}$. At low temperatures Anderson localisation occurs in many metals and semiconductors where variable range hopping of the charge carriers is observed. In the variable range hopping mechanism, a localised electron can then only move from one localised site to another by phonon assisted hopping, which is a combined thermally active quantum tunnelling process. An electron will only tunnel to another site if the thermal activation energy required for the hop is reduced. In this case the resistivity, $\rho$, can be modelled as $\rho = \rho_0 \exp(T_0/T)^{1/4}$. $T_0$ represents the localisation temperature. $T_0$ depends on the radius of localised states $a$ (also called the localisation length), and the density of states at the Fermi level, $N(E_F)$ so that $T_0 = \lambda/[k_B N(E_F) a^3]$; $\lambda$ is a dimensionless constant.[27, 28] Three-dimensional variable-range hopping of the electrons is also observed for NdMnAsO$_{0.95}$F$_{0.05}$ below 75 K [11]. The resistivity data for PrMnAsO$_{0.95}$F$_{0.05}$ in Figure 7 is fit to the Mott three-dimensional variable-range hopping (VRH) equation. A subtle electronic transition is evidenced at 44 K (Fig.7) as a change of slope is observed in the ln($\rho$) versus $T^{-0.25}$ plot. This transition is not coincident with the structural transition at 35 K. The change of slope observed in the ln($\rho$) versus $T^{-0.25}$ plot (Fig. 7) suggests a crossover between two three-dimensional VRH states with different values of the localisation temperature $T_0$ ($T_0 = 6.16(1) \times 10^5$ K and $2.08(1) \times 10^5$ K for the high and low temperature electronic states respectively). This reveals that the material adopts a more disordered electronic state at higher temperatures as $T_0$, which represents the degree of electronic disorder, is larger. $T_0$ is large for the LnMnAsO$_{1-x}$F$_x$ (Ln = Pr, Nd) oxypnictides reported so far which suggests that electronic disorder is considerable.[28] In this case conduction can only take place by an exchange transition to the final state so that the electron reaches its final state by an indirect, rather than a single, transition. [28] The electron percolates through the network in a set of varied jumps, rather than a single jump, and the additional energy is supplied to the electron by a multiphonon process. The double slope behaviour observed for $PrMnAsO_{0.95}F_{0.05}$ has been reported in the insulating phase of high temperature superconducting cuprates [28] and also in $SrFeO_{3-\delta}$ [29] and is due to multiphonon assisted hopping in different temperature regimes. Below 12 K a further electronic transition is evidenced for $PrMnAsO_{0.95}F_{0.05}$ so that variable range hopping of the electrons is no longer observed. The variation of resistivity with temperature is displayed in inset a, Figure 7 and the resistivity appears to reach a maximum at 4 K. The origin of this transition is currently unknown.

Figure 8 shows the variation of the magnetoresistance with temperature for both $PrMnAsO_{0.95}F_{0.05}$ and $NdMnAsO_{0.95}F_{0.05}$ recorded in a 7 T magnetic field. The MR of $PrMnAsO_{0.95}F_{0.05}$ is very different to that reported for $NdMnAsO_{1-x}F_x$ ($x \geq 0.05$) and there is no evidence of CMR down to 4 K. Instead, upon cooling below the structural transition at ~ 35 K a sizeable negative MR is observed ($MR_{7T}$ (12 K) = -13.4 %). Below 12 K the magnitude of the –MR rapidly decreases so that by 4 K the $MR_{7T}$ = -0.9 %. Below 12 K a further electronic transition is evidenced (as described above) so that VRH of the charge carriers is no longer observed and demonstrates that the MR is sensitive to the conduction mechanism. There is no evidence of MR above 35 K.

**Discussion**

The magnetic properties of $NdMnAsO_{0.95}F_{0.05}$ and $PrMnAsO_{0.95}F_{0.05}$ exhibit clear differences. In both materials there is significant magnetic coupling between the lanthanide and Mn ions. In $PrMnAsO_{0.95}F_{0.05}$ the Mn spin reorientation transition and subsequent induced antiferromagnetic order of $Pr^{3+}$ occurs below ~ 220 K and over a much wider temperature range (~ 80 K compared to 3 K in $NdMnAsO_{0.95}F_{0.05}$). The Nd antiferromagnetic transition and subsequent spin reorientation of Mn moments in $NdMnAsO_{0.95}F_{0.05}$ also occurs at a much lower temperature (23 K). [11] Density functional calculations have shown that the spin reorientation transitions observed in LnMnAsO (Ln = lanthanide) are controlled by the strength of the Dzyaloshinskii-Moriya (DM) and biquadratic (BQ) exchanges between $Ln^{3+}$ and Mn. [30] Below the SR transition the Mn and $Pr^{3+}$ spins are collinear so that the BQ interaction dominates over the DM interaction. The higher spin reorientation transition temperature observed in $PrMnAsO_{0.95}F_{0.05}$ compared to $NdMnAsO_{0.95}F_{0.05}$ therefore demonstrates a much stronger BQ interaction between Mn and $Ln^{3+}$ in $PrMnAsO_{0.95}F_{0.05}$. [30]

Figure 8 shows that the MR properties of $PrMnAsO_{0.95}F_{0.05}$ are very different to $NdMnAsO_{0.95}F_{0.05}$ in which colossal magnetoresistance is observed at low temperature ($MR_{9T}(3K)$) = -95%). Both materials are semiconducting with Arrhenius behaviour at high temperature and variable range hoping of the charge carriers below ~ 80 K. However, surprisingly, the CMR observed for $NdMnAsO_{1-x}F_x$ is absent in $PrMnAsO_{0.95}F_{0.05}$. -MR is observed below ~ 75 K in $NdMnAsO_{0.95}F_{0.05}$ but at $T_{SR}$ the magnitude of -MR increases sharply (Fig. 8). At $T_{SR}$, in $NdMnAsO_{1-x}F_x$, the spin reorientation of the Mn spins from aligning along $c$ to aligning parallel to $a$ precipitates an electronic transition to Efros Shklovskii (ES) VRH. This signifies that the reorientation of Mn spins into the basal plane at 20 K results in enhanced Coulomb correlations between localized electrons, [11] which results in much higher resistivity below $T_{SR}$. A variable field neutron diffraction study has shown that upon applying a magnetic field there is a second order phase transition from antiferromagnetic to paramagnetic order of both the Nd and Mn spins. [11] The CMR then arises as a result of a second order phase transition from an insulating antiferromagnet to a semiconducting paramagnet upon applying a magnetic field so that the electron correlations are diminished in field. [11] The MR at a given field, H, is related to the magnitude of the antiferromagnetically ordered Mn moment, so that $-MR = \left(\frac{\Delta M}{C}\right)^{1/2}$ where $\Delta M = M(0) - M(H)$, where M(H) is the $Mn^{2+}$ moment in field, M(0) is the $Mn^{2+}$ moment when $\mu_0 H = 0$ T and C is a constant (0.4 $\mu_B$ at 4 K) which equates to the moment reduction that would theoretically result in a -MR of 100%.

The temperature variation of the MR of PrMnAsO$_{0.95}$F$_{0.05}$ is very different (Fig. 8) and a sizeable – MR is only observed below $T_s$. Below $T_s$ multipolar order of the Pr 4f ground state is observed so that the sizeable –MR observed most likely arises due to a reduction in magnetic and/or multipolar scattering upon application of a magnetic field. The absence of CMR in PrMnAsO$_{0.95}$F$_{0.05}$ is perhaps surprising and most likely suggests that the phase transition from antiferromagnetic to paramagnetic order of both the Ln and Mn spins observed from variable field neutron diffraction studies of NdMnAsO$_{0.95}$F$_{0.05}$ (which results in the CMR) is not evident in PrMnAsO$_{0.95}$F$_{0.05}$. It has previously been proposed that the antiferromagnetic to paramagnetic phase transition in NdMnAsO$_{1-x}$F$_x$ arises as a result of a linear coupling between a hidden order parameter and the Mn and/or Nd antiferromagnetic moments. [11] Such behaviour has also been reported for the heavy fermion material URu$_2$Si$_2$. [31] The absence of CMR in PrMnAsO$_{0.95}$F$_{0.05}$ would suggest that the hidden order parameter is no longer present, and most likely originates from the Nd$^{3+}$ cation. The hidden order parameter could, for example, originate from multipolar ordering of the Nd$^{3+}$ 4f ground state. Further studies are warranted to gain a better understanding of the complex physics reported for NdMnAsO$_{0.95}$F$_{0.05}$. For PrMnAsO$_{0.95}$F$_{0.05}$, there is also no evidence of an Efros Shklovskii VRH transition ($T_{ES}$) down to 4 K. Hence electron correlations are weaker in PrMnAsO$_{0.95}$F$_{0.05}$ so that the contribution to the CMR observed in NdMnAsO$_{1-x}$F$_x$ below the simultaneous spin reorientation and ES transition is absent. The weaker electron correlations in PrMnAsO$_{0.95}$F$_{0.05}$ could be a result of the structural change from tetragonal to orthorhombic symmetry below 35 K which results in a longer Mn-As bond length along *b* (as the Mn moments align parallel to *b* below $T_s$). In comparison at 10 K the Mn-As bond length is 2.5424(5) in NdMnAsO$_{0.95}$F$_{0.05}$, compared to 2.556(1) in PrMnAsO$_{0.95}$F$_{0.05}$.

We note that it is unlikely that the difference in the MR behaviour of PrMnAsO$_{0.95}$F$_{0.05}$ and NdMnAsO$_{0.95}$F$_{0.05}$ is a result of small differences in O:F stoichiometry. The temperature variation of the MR of NdMnAsO$_{0.98}$F$_{0.02}$ is very similar to NdMnAsO$_{0.95}$F$_{0.05}$, where a large drop in MR is observed at $T_{SR}$ (Fig. 8, inset). However, the overall MR is much smaller (MR$_{7T}$ (10 K) ~ -29%).

## Conclusions

Similar to NdMnAsO$_{0.95}$F$_{0.05}$, PrMnAsO$_{0.95}$F$_{0.05}$ is a semiconductor with magnetic coupling between the rare earth and Mn magnetic spins. Antiferromagnetic order of the Mn spins with moments aligned parallel to *c* is observed below 340 K for PrMnAsO$_{0.95}$F$_{0.05}$. Below 220 K the Mn spins begin to reorient into the basal plane, which subsequently induces antiferromagnetic order of the Pr spins at 180 K. However, unlike NdMnAsO$_{0.95}$F$_{0.05}$, a tetragonal to orthorhombic structural transition is detected below 35 K, which is driven by multipolar ordering of the Pr 4f ground state and is accompanied by a sizeable negative magnetoresistance below $T_s$. The absence of CMR in PrMnAsO$_{0.95}$F$_{0.05}$ may be attributed to several factors, including the absence of Efros Shklovskii VRH as a result of weakened electron correlations due to longer Mn-As bond lengths below $T_S$. These results demonstrate that changing the lanthanide in 1111 *Ln*MnAsO$_{0.95}$F$_{0.05}$ strongly impacts the electronic and magnetic properties, so that multiple mechanisms of MR are possible in Mn oxypnictides.

## Acknowledgements

This research is supported by the EPSRC (research grant EP/L002493/1). We also acknowledge STFC-GB for provision of beamtime at ILL and ESRF.

## Author Information

a.c.mclaughlin@abdn.ac.uk (corresponding author)

Table.1 Refined cell parameters, agreement factors, atomic parameters and selected bond lengths and angles for PrMnAsO$_{0.95}$F$_{0.5}$ from Rietveld fits against ID31 synchrotron powder X-ray diffraction data at various temperatures. Pr and As are at 2c (¼, ¼, z), Mn at 2b (¾, ¼, z) and O, F at 2a (¾, ¼, z). A structural transition from tetragonal *P*4*/nmm* to orthorhombic *Pmmn* symmetry is apparent below 35 K.

| Atom | Occupancy | | Temperature (K) | | | | | | |
|---|---|---|---|---|---|---|---|---|---|
| | | | 10 | 15 | 20 | 25 | 30 | 35 | 40 |
| Pr | 1.00 | z | 0.13150(3) | 0.13146(3) | 0.13148(3) | 0.13147(3) | 0.13143(3) | 0.13143(3) | 0.13143(3) |
| | | $U_{iso}$ (Å$^2$) | 0.00112(4) | 0.00112(4) | 0.00102(4) | 0.00116(4) | 0.00124(4) | 0.00121(4) | 0.00122(4) |
| Mn | 1.00 | z | 0.5014(3) | 0.5013(3) | 0.5006(3) | 0.5009(3) | 0.5010(4) | 0.5 | 0.5 |
| | | $U_{iso}$ (Å$^2$) | 0.0018(1) | 0.0020(1) | 0.0020(1) | 0.00197(12) | 0.0022(1) | 0.0019(1) | 0.0020(1) |
| As | 1.00 | z | 0.67299(5) | 0.67297(5) | 0.67300(5) | 0.67298(5) | 0.67289(5) | 0.67288(5) | 0.67292(6) |
| | | $U_{iso}$ (Å$^2$) | 0.00149(9) | 0.00155(8) | 0.00150(8) | 0.00160(8) | 0.00166(8) | 0.00159(8) | 0.00165(9) |
| O/F | 0.95/0.05 | z | 0.008(1) | 0.0084(9) | 0.0097(9) | 0.0092(9) | 0.009(1) | 0 | 0 |
| | | $U_{iso}$ (Å$^2$) | 0.0001(7) | 0.0001(6) | 0.0005(6) | 0.0007(6) | 0.0009(7) | 0.0028(5) | 0.0026(5) |

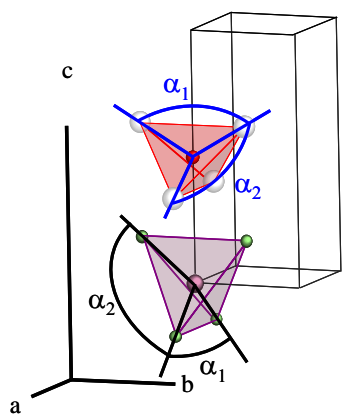

| | | | | | | | |
|---|---|---|---|---|---|---|---|
| $a$ (Å) | 4.05896(1) | 4.059057(9) | 4.059149(9) | 4.059300(9) | 4.05953(1) | 4.060639(6) | 4.060731(6) |
| $b$ (Å) | 4.06201(1) | 4.061925(9) | 4.061887(9) | 4.061805(9) | 4.06161(1) | - | - |
| $c$ (Å) | 8.89399(2) | 8.89444(1) | 8.89482(1) | 8.89513(1) | 8.89546(1) | 8.89557(2) | 8.89590(2) |
| $\chi^2$ (%) | 2.785 | 3.926 | 3.597 | 3.355 | 3.133 | 3.108 | 2.967 |
| $R_{WP}$ (%) | 12.14 | 11.46 | 11.65 | 11.72 | 11.80 | 12.24 | 12.47 |
| $R_P$ (%) | 8.43 | 8.03 | 8.19 | 8.42 | 8.23 | 8.63 | 8.79 |
| Pr-O/F (Å) | 2.378(5) | 2.381(5) | 2.388(4) | 2.386(5) | 2.385(5) | 2.3429(1) | 2.3429(1) |
| | 2.309(5) | 2.306(4) | 2.301(4) | 2.303(4) | 2.303(4) | - | - |
| Mn-As (Å) | 2.556(1) | 2.555(1) | 2.551(1) | 2.553(1) | 2.553(2) | 2.5470(3) | 2.5473(3) |
| | 2.539(1) | 2.540(1) | 2.544(1) | 2.542(1) | 2.542(2) | - | - |
| Mn-Mn (Å) | 2.87131(5) | 2.87129(4) | 2.87124(2) | 2.87129(3) | 2.87131(4) | 2.87131(4) | 2.87137(2) |
| Pr-As (Å) | 3.3567(3) | 3.3570(3) | 3.3568(3) | 3.3570(3) | 3.3577(3) | 3.3578(3) | 3.3577(3) |
| $\alpha_1$ Pr-O/F-Pr (°) | 123.0(4) | 123.3(4) | 123.8(4) | 123.6(4) | 123.6(4) | 120.13(1) | 120.13(1) |

| | | 117.3(4) | 117.0(4) | 116.5(3) | 116.7(4) | 116.8(4) | - | - |
|---|---|---|---|---|---|---|---|---|
| | $\alpha_2$ Pr-O/F-Pr (°) | 104.37(1) | 104.36(1) | 104.34(1) | 104.35(1) | 104.34(1) | 104.419(5) | 104.420(5) |
| | $\alpha_1$ As-Mn-As (°) | 111.39(1) | 111.391(9) | 111.399(9) | 111.396(9) | 111.381(9) | 111.38(1) | 111.39(1) |
| | $\alpha_2$ As-Mn-As (°) | 106.1(1) | 106.1(1) | 105.8(1) | 105.9(1) | 106.0(1) | 105.72(1) | 105.70(1) |
| | | 105.3(1) | 105.3(1) | 105.5(1) | 105.4(1) | 105.4(1) | - | - |
| | MnAs Layer (Å) | 3.0522(2) | 3.0538(1) | 3.0669(1) | 3.0613(1) | 3.0581(1) | 3.0757(2) | 3.0766(2) |
| | Pr(O/F) Layer (Å) | 2.2039(2) | 2.1891(1) | 2.1664(1) | 2.1752(1) | 2.1781(1) | 2.338(2) | 2.3384(2) |

| Atom | Occupancy | | Temperature (K) | | | | | | |
|---|---|---|---|---|---|---|---|---|---|
| | | | 70 | 100 | 130 | 160 | 200 | 250 | 290 |
| Pr | 0.995 | z | 0.13143(3) | 0.13135(3) | 0.13139(3) | 0.13135(3) | 0.13124(3) | 0.13123(3) | 0.13119(3) |
| | | $U_{iso}$ (Å$^2$) | 0.00174(5) | 0.00199(4) | 0.00295(5) | 0.00291(4) | 0.00359(5) | 0.00445(5) | 0.00501(5) |
| Mn | 1.00 | z | 0.5 | 0.5 | 0.5 | 0.5 | 0.5 | 0.5 | 0.5 |
| | | $U_{iso}$ (Å$^2$) | 0.0027(1) | 0.0030(1) | 0.0042(1) | 0.0043(1) | 0.0052(1) | 0.0066(1) | 0.0075(1) |
| As | 1.00 | z | 0.67291(6) | 0.67291(5) | 0.67280(6) | 0.67277(6) | 0.67287(6) | 0.67301(6) | 0.67295(6) |
| | | $U_{iso}$ (Å$^2$) | 0.00227(9) | 0.00264(8) | 0.0038(1) | 0.00398(9) | 0.0048(1) | 0.0060(1) | 0.0067(1) |
| O/F | 0.95/0.05 | z | 0 | 0 | 0 | 0 | 0 | 0 | 0 |
| | | $U_{iso}$ (Å$^2$) | 0.0034(6) | 0.0034(5) | 0.0059(7) | 0.0042(6) | 0.0039(6) | 0.0045(6) | 0.0051(6) |
| | | a (Å) | 4.061816(6) | 4.062342(6) | 4.062877(6) | 4.063972(6) | 4.065303(6) | 4.067319(6) | 4.068873(6) |
| | | c (Å) | 8.89837(2) | 8.90023(1) | 8.90198(2) | 8.90543(1) | 8.90982(2) | 8.91620(1) | 8.92088(1) |
| | | $\chi^2$ (%) | 4.428 | 3.456 | 3.677 | 2.969 | 2.792 | 3.876 | 6.166 |
| | | $R_{WP}$ (%) | 12.73 | 11.04 | 12.53 | 11.55 | 11.86 | 11.24 | 10.94 |
| | | $R_P$ (%) | 9.59 | 8.13 | 9.01 | 8.19 | 8.39 | 8.14 | 7.65 |
| | | Pr-O/F (Å) | 2.3436(1) | 2.3436(1) | 2.3441(1) | 2.3446(1) | 2.3450(1) | 2.3463(1) | 2.3470(1) |
| | | Mn-As (Å) | 2.5479(3) | 2.5483(3) | 2.5482(4) | 2.5488(3) | 2.5503(3) | 2.5525(3) | 2.5533(3) |

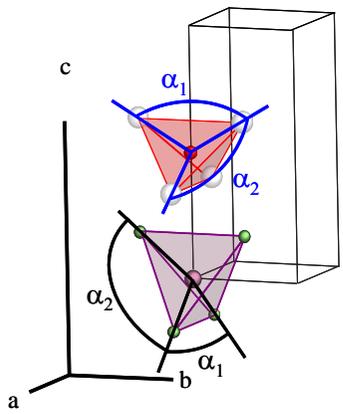

| | | | | | | | |
|---|---|---|---|---|---|---|---|
| Mn-Mn (Å) | 2.87214(2) | 2.87251(3) | 2.87289(1) | 2.87366(1) | 2.87460(2) | 2.87603(1) | 2.87713(4) |
| Pr-As (Å) | 3.3586(3) | 3.3595(3) | 3.3603(4) | 3.3617(3) | 3.3630(4) | 3.3642(3) | 3.3660(3) |
| $\alpha_1$ Pr-O-Pr (°) | 120.13(1) | 120.15(1) | 120.14(1) | 120.15(1) | 120.18(1) | 120.17(1) | 120.18(1) |
| $\alpha_2$ Pr-O-Pr (°) | 104.420(6) | 104.409(5) | 104.416(6) | 104.412(6) | 104.397(6) | 104.401(6) | 104.398(6) |
| $\alpha_1$ As-Mn-As (°) | 111.39(1) | 111.388(9) | 111.37(1) | 111.37(1) | 111.39(1) | 111.42(1) | 111.42(1) |
| $\alpha_2$ As-Mn-As (°) | 105.70(2) | 105.70(1) | 105.73(2) | 105.73(2) | 105.69(2) | 105.64(2) | 105.65(2) |
| MnAs Layer (Å) | 3.0772(2) | 3.0779(1) | 3.0765(2) | 3.0772(1) | 3.0805(2) | 3.0852(1) | 3.0857(1) |
| Pr(O/F) Layer (Å) | 2.3390(2) | 2.3381(1) | 2.3393(2) | 2.3395(1) | 2.3387(2) | 2.3401(1) | 2.3407(1) |

**Figure Captions**

Figure 1 Rietveld refinement fit to the 10 K ID31 synchrotron X-ray powder diffraction pattern of PrMnAsO$_{0.95}$F$_{0.05}$. The inset shows the orthorhombic peak splitting as a function of temperature.

Figure 2 Temperature variation of the *a* and *b* cell parameters of PrMnAsO$_{0.95}$F$_{0.05}$ with temperature, evidencing the structural transition below 35 K. The inset shows the variation of the *c* cell parameter with temperature. The temperature at which the Mn spins are fully aligned in the basal plane, $T_{SR2}$, is indicated on both figures by an arrow.

Figure 3 Variation of the Pr-O/F and Mn-As bond lengths with temperature. The inset shows the temperature variation of the As-Mn-As bond angle ($\alpha_1$). The structural anomaly at $T_{SR2}$ is indicated by an arrow.

Figure 4 A portion of the D20 neutron diffraction pattern for PrMnAsO$_{0.95}$F$_{0.05}$ showing the magnetic diffraction peaks at several temperatures. The Rietveld fit to the data is shown. Diffuse scattering characteristic of short range magnetic order is still evident above $T_{Mn}$ as indicated by the arrow. The inset shows the magnetic structure below $T_{Pr}$. The small black spheres are Mn and the larger grey spheres are Pr.

Figure 5 The temperature variation of the refined Mn and Pr moments for PrMnAsO$_{0.95}$F$_{0.05}$. The shaded region shows the spin reorientation of the Mn spins from parallel to *c* into the basal plane between $T_{SR1}$ and $T_{SR2}$.

Figure 6 Dc-magnetic susceptibility of PrMnAsO$_{0.95}$F$_{0.05}$ measured as a function of temperature in a 1000 Oe magnetic field. The inset shows a section of the variation of the inverse susceptibility with temperature which evidences antiferromagnetic order of the Mn spins below 340 K. The bold black line is a linear fit to the inverse susceptibility data and there is a clear deviation at $T_{Mn}$.

Figure 7 Plot of ln(resistivity) against $T^{-1/4}$ for PrMnAsO$_{0.95}$F$_{0.05}$ evidencing a subtle electronic transition at 44 K. The resistivity can be fit to the Mott VRH equation $\rho = \rho_0 \exp(T_0/T)^{1/4}$ over the whole temperature range shown. The insets show the variation of resistivity for PrMnAsO$_{0.95}$F$_{0.05}$

(inset a) and variation of ln(ρ) with inverse temperature showing that PrMnAsO$_{0.95}$F$_{0.05}$ exhibits Arrhenius behaviour between 95 K – 220 K (inset b).

Figure 8 The temperature variation of the MR$_{7T}$ for PrMnAsO$_{0.95}$F$_{0.05}$ and NdMnAsO$_{0.95}$F$_{0.05}$ demonstrating the absence of CMR in PrMnAsO$_{0.95}$F$_{0.05}$. The structural transition (T$_s$) observed at ~ 35 K for PrMnAsO$_{0.95}$F$_{0.05}$ and the spin reorientation transition (T$_{SR}$) observed at 20 K for NdMnAsO$_{0.95}$F$_{0.05}$ are indicated. The inset shows the temperature variation of MR$_{7T}$ for NdMnAsO$_{0.98}$F$_{0.02}$.

Figure 1

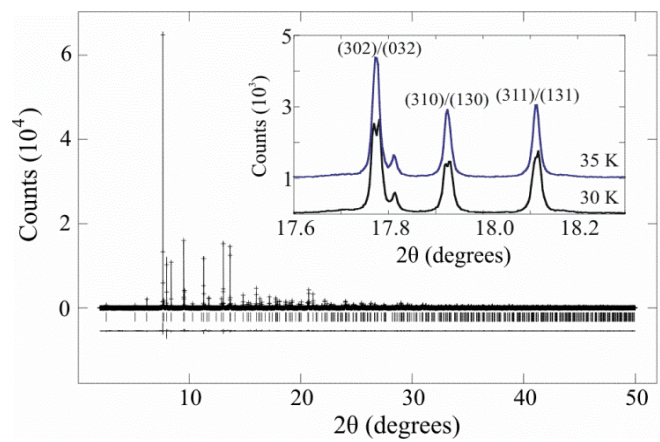

Figure 2

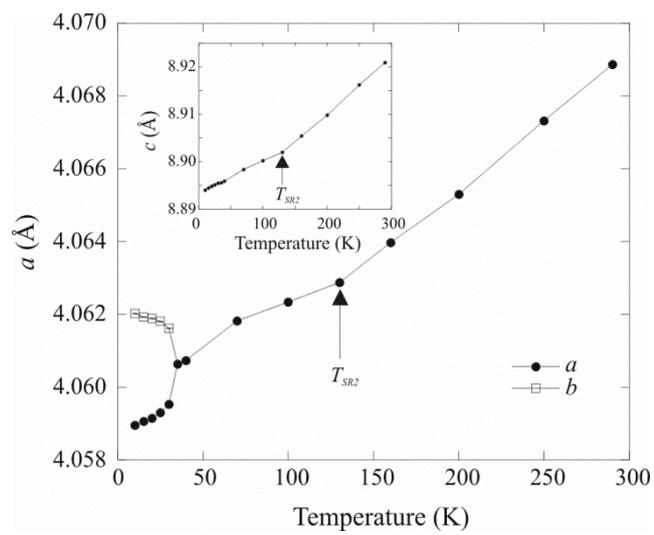

Figure 3

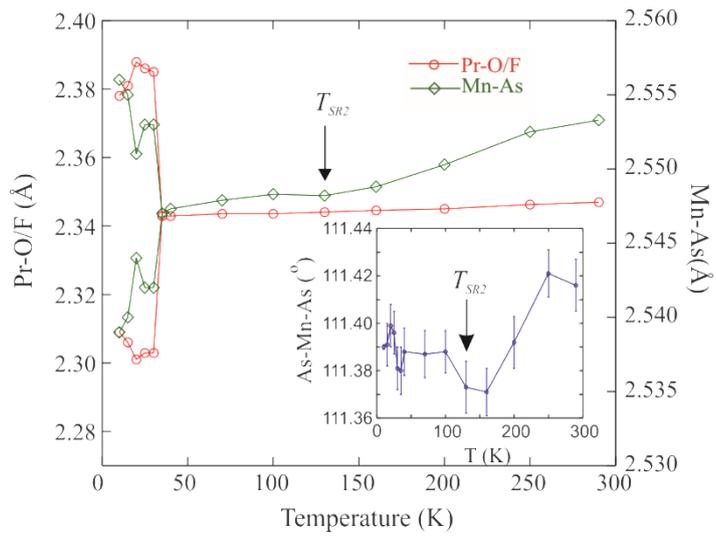

Figure 4

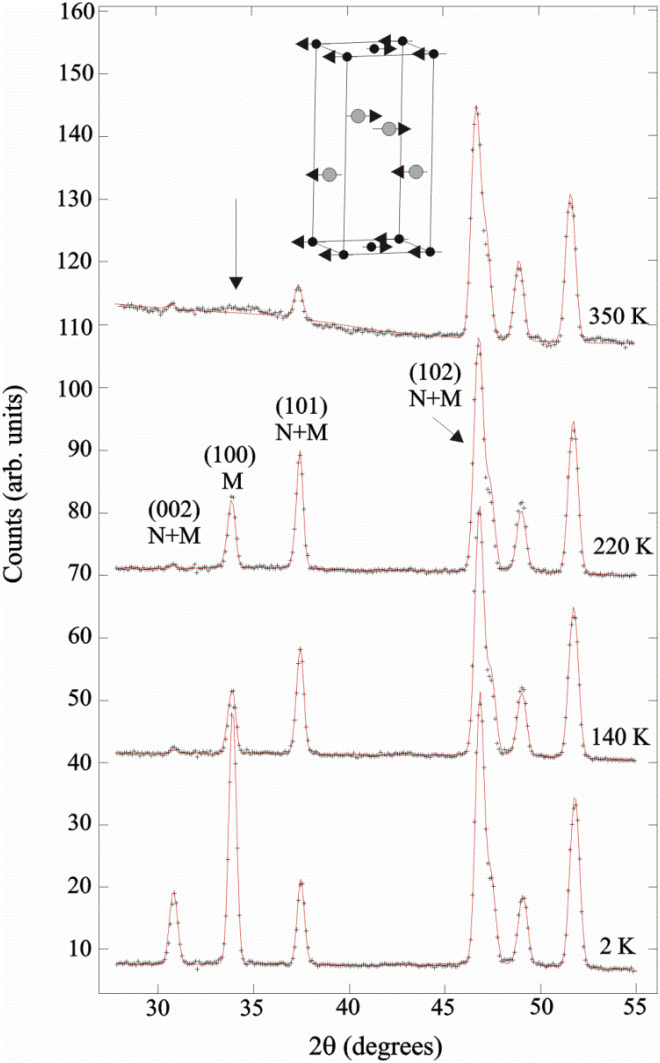

Figure 5

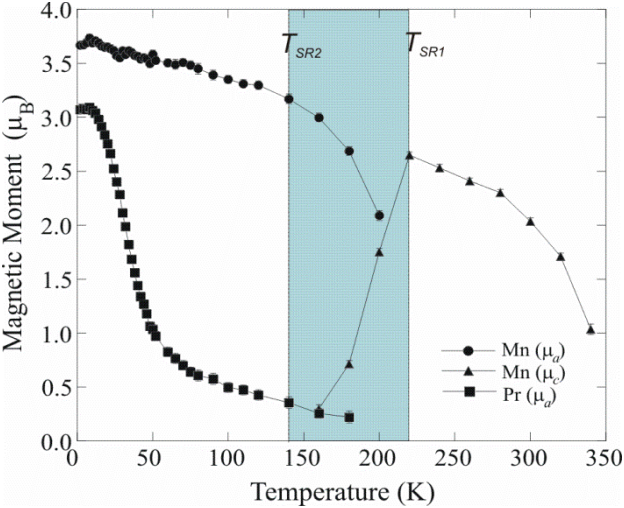

Figure 6

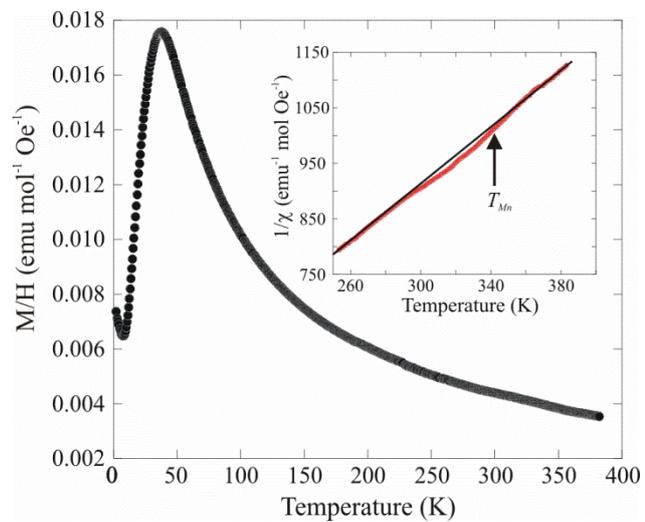

Figure 7

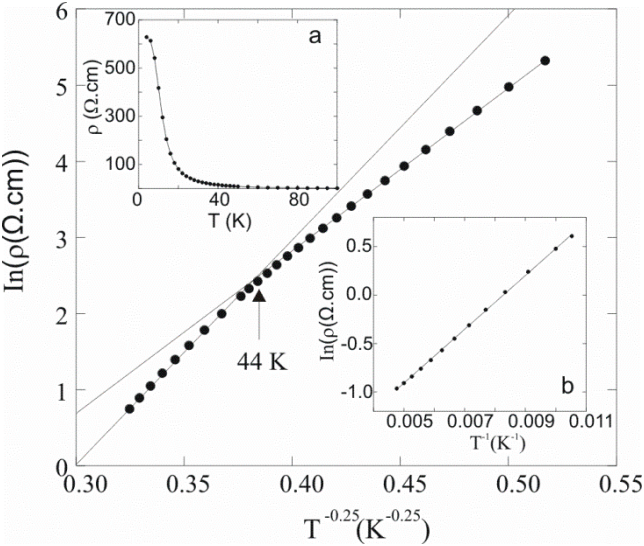

Figure 8

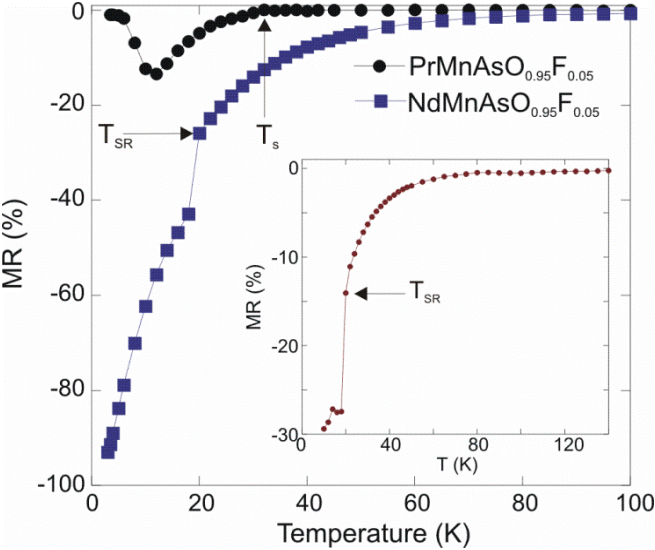

## For Table of Contents Only

The colossal magnetoresistance observed for NdMnAsO$_{0.95}$F$_{0.05}$ is absent in PrMnAsO$_{0.95}$F$_{0.05}$. Instead a sizeable negative magnetoresistance is observed below a structural transition at 35 K.

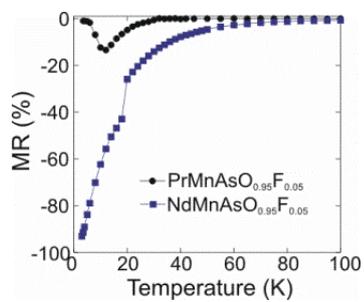